\documentclass[twocolumn,showpacs,amsmath,amssymb]{revtex4}
\usepackage{graphicx}
\usepackage{dcolumn}
\usepackage{bm}

\newcommand{\D}{{\rm d}}

\newcommand{\Del}{\Delta}

\newcommand{\eq}{{\rm eq}}

\newcommand{\Hz}{H_{z}}
\newcommand{\ho}{h_{0}}

\newcommand{\iu}{{\rm i}}

\newcommand{\K}{D}

\newcommand{\w}{\omega}

\newcommand{\el}{\varepsilon} 

\newcommand{\Xone}{\chi_{1}}
\newcommand{\Xtwo}{\chi_{2}}
\newcommand{\Xnl}{\chi_{3}}

\begin{document}



\title{
Non-linear response of single-molecule magnets:
\\
field-tuned quantum-to-classical crossovers
}

\author{
R.~L\'{o}pez-Ruiz, F. Luis,$^{\ast}$
A. Mill\'an, C. Rillo,
D. Zueco and J.~L. Garc\'{\i}a-Palacios
}
\affiliation{
Instituto de Ciencia de Materiales de Arag\'on y
Dep.\ de F\'{\i}sica de la Materia Condensada,
C.S.I.C. -- Universidad de Zaragoza, E-50009 Zaragoza, Spain
}


\date{\today}

\begin{abstract}
Quantum nanomagnets can show a field dependence of the relaxation time
very different from their classical counterparts, due to resonant
tunneling via excited states (near the anisotropy barrier top).
The relaxation time then shows minima at the resonant fields
$H_{n}\propto n\K$ at which the levels at both sides of the barrier
become degenerate ($\K$ is the anisotropy constant).
We showed that in Mn$_{12}$, near zero field, this yields a
contribution to the {\em nonlinear\/} susceptibility that makes it
qualitatively different from the classical curves [Phys.\ Rev.\ B {\bf
72}, 224433 (2005)].
Here we extend the experimental study to finite dc fields showing how
the bias can trigger the system to display those quantum nonlinear
responses, near the resonant fields, while recovering an
classical-like behaviour for fields between them.
The analysis of the experiments is done with heuristic expressions
derived from simple balance equations and calculations with a
Pauli-type quantum master equation.
\end{abstract}

\pacs{75.50.Xx, 75.50.Tt, 75.45.+j, 75.40.Gb}

\keywords{
Single-molecule magnets, nonlinear susceptibilities, magnetic
relaxation, quantum tunneling, quantum dissipative systems.
}

\maketitle



\section{
Introduction
}

Single-molecule magnets are metal-organic clusters containing a
magnetic core surrounded by a shell of organic ligands which isolates
the clusters from one another (see, e.g., Refs.~\cite{smm}).
The most studied is Mn$_{12}$, whose core contains eight Mn$^{3+}$ and
four Mn$^{4+}$ ions strongly coupled via super-exchange
interactions.
This gives a ground state spin $S=10$, while the large Jahn-Teller
distortion on the Mn$^{3+}$ sites leads to a strong uniaxial
anisotropy.
The energy levels have a bistable structure $\el_{m}\sim-\K\,m^{2}$ (at
zero field) with an energy barrier $U=\el_{0}-\el_{S}\simeq70$\,K to
be overcome for the spin reversal.
At low temperatures, these systems show the typical behaviors of
superparamagnets, such as blocking or hysteresis, yet at a much
smaller scale of size.
In addition, they form molecular crystals in which all molecules are
nearly identical and, in the case of Mn$_{12}$ acetate, have their
anisotropy axes $\boldsymbol{z}$ parallel to the crystallographic
$\boldsymbol{c}$ axis.

These properties make molecular magnets model systems to investigate
whether quantum phenomena, like tunneling, survive in mesoscopic
systems \cite{leg2002}.
As is well-known, tunneling probabilities decrease exponentially with
the height of the barrier to be tunneled through (height that grows
with the system size) \cite{magtun}.
At the same time, external perturbations can induce decoherence that
degrades the quantum behavior \cite{zur91,kiejoo99}.
Furthermore, an external magnetic field $\Hz$ detunes energetically
the initial and final states for tunneling (i.e., those having $+m$
and $-(m+n)$ spin projections along $\boldsymbol{z}$).
Actually, many experiments have shown that tunneling takes place at
those fields where states of opposite orientation are degenerate,
$H_{n}\simeq n\,H_{1}$ ($n=0,1,2,\dots$ with $H_{1}=2g\mu_{\rm
B}D\simeq4200$\,Oe in Mn$_{12}$), whereas it is suppressed for
intermediate fields \cite{frietal96,heratal96,thoetal96}.

Such a {\em resonant} tunneling enables the spins to approach faster
their equilibrium state, giving rise to steps in the hysteresis loops
around $\Hz=H_{n}$ \cite{wer2001} and to maxima in the linear
dynamical susceptibility $\Xone$.
In our previous work \cite{luietal2004a,lopetal2005}, we found that
resonant tunneling at $\Hz=0$ induces an extra contribution to the
nonlinear response $\Xnl$, making it larger (in magnitude) than the
equilibrium one and having peaks reversed with respect to the
classical predictions \cite{garsve2000,gargar2004}.
In Refs.~\cite{luietal2004a,lopetal2005} the dependence of the
response on temperature, frequency, and orientation at zero field was
studied.
In this article we show how the tunneling contribution to the
nonlinear response can be switched on and off by varying an external
field, tuning and breaking successively the resonances.


\section{
Experimental details
}
\label{sec:experimentalia}

Single crystals of Mn$_{12}$ acetate were grown following the same
procedure described in \cite{lopetal2005}.
In order to increase the signal, the magnetic measurements were done
on a collection of oriented and glued crystals.
In our previous experiments \cite{luietal2004a,lopetal2005} we
extracted the zero-field $\Xnl(\w)$ by fitting the dc field dependent
$\chi(\w)$ to a parabola.
Clearly, this method is not applicable to study how $\Xnl$ depends on
the external magnetic field itself.
For this reason, in the present experiments we resorted to the more
traditional method of measuring non-linear susceptibilities by
detecting the different harmonics $\Xtwo(2\w)$ and $\Xnl(3\w)$ of the
response.
In the absence of bias field one had $\Xtwo\equiv0$ (see below).
A nonzero $\Hz$, however, makes $\Xtwo$ the leading nonlinear term,
and we will mainly focus on it.
\begin{figure}[t]
\resizebox{8.cm}{!}{\includegraphics{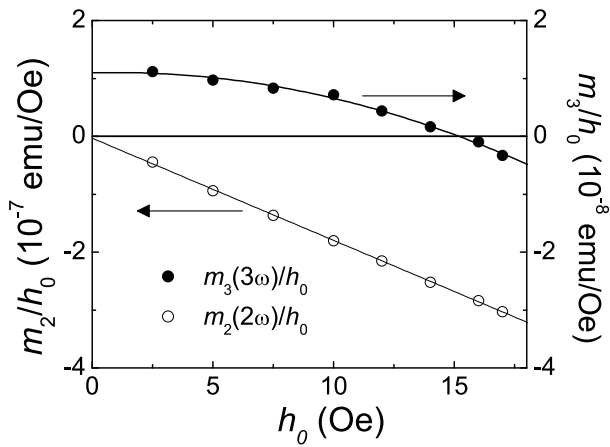}}
\caption{
Illustration of the method employed for measuring the nonlinear
susceptibilities (here $T=8$\,K, $\Hz=100$\,Oe, and $\w/2\pi=2$\,kHz).
$\Xtwo(2\w)$ and $\Xnl(3\w)$ are obtained, respectively, from the
slope and the quadratic coefficient of the second $m_{2}(2\w)/\ho$ and
third $m_{3}(3\w)/\ho$ harmonics of the output signal, measured as a
function of the ac field amplitude $\ho$.
}
\label{fig1}
\end{figure}

The fields, both dc and ac, were applied {\em parallel\/} to the
common anisotropy direction $\boldsymbol{z}$ of the clusters.
We employed the susceptibility option of a commercial multipurpose
measuring platform (PPMS) which uses a conventional inductive method.
It enables applying ac fields of amplitude $\ho\leq17$\,Oe, and to
selectively detect several harmonics of the exciting frequency
$\w/2\pi<10$\,kHz.
To separate the intrinsic nonlinear response of the sample from the
possible contamination due to non-perfect harmonicity of the exciting
ac coil, we measured the output signals $m_{2}(2\w)/\ho$ and
$m_{3}(3\w)/\ho$ at several $\ho$.
This gives the sought-for intrinsic contributions $\Xtwo$ and $\Xnl$
as the terms proportional to $\ho$ and $\ho^{2}$ respectively.
An example of this procedure is shown in Fig.~\ref{fig1}.
\begin{figure}
\resizebox{8.5cm}{!}{\includegraphics{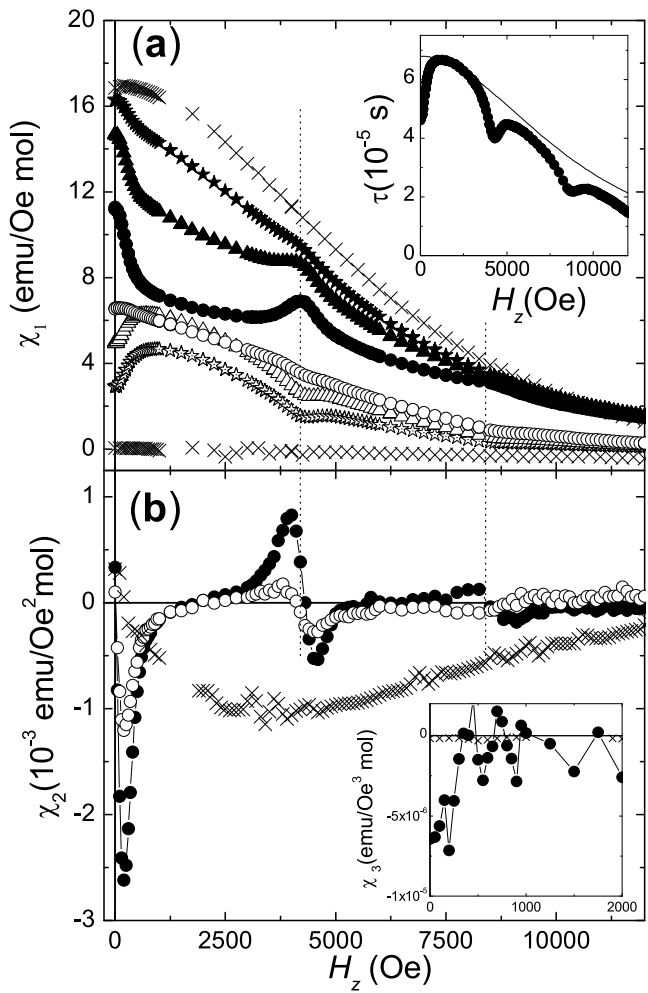}}
\caption{
Upper panel: linear susceptibility of Mn$_{12}$ measured at $T=8$\,K
versus a magnetic field applied parallel to the anisotropy axis.
$\times$, $\w/2\pi=1$\,Hz ($\sim$equilibrium); $\star$, 500\,Hz;
$\triangle$, 1\,kHz; $\bullet$, 2\,kHz.
Solid symbols, real parts; open symbols, imaginary parts.
The inset shows the relaxation time $\tau$, as obtained from
$\Xone''/\w\Xone'$ [see Eq.~(\ref{X1:debye})] as well as calculated
for classical spins (line) \cite{bro63}.
Lower panel, second harmonic susceptibility measured at the same
temperature.
$\bullet$ and $\circ$, $\Xtwo'(2\w)$ and $\Xtwo''(2\w)$ at 2\,kHz;
$\times$, equilibrium $\Xtwo^{\eq} = (\D\Xone^{\eq} /\D\Hz)/2$.
Inset: $\bullet$, $\Xnl'(3\w)$ at 2\,kHz; $\times$,
$\Xnl^{\eq}=(\D^{2}\Xone^{\eq}/\D\Hz^{2})/6$.
The dotted vertical lines mark the resonant fields
$H_{1}\simeq4200$\,Oe and $H_{2}=2\,H_{1}$.
}
\label{fig2}
\end{figure}


\section{
Results and modelization
}
\label{results}

The so-measured linear and nonlinear susceptibilities of Mn$_{12}$ at
$T=8$\,K are shown in Fig.~\ref{fig2}.
The frequency-dependent $\Xone$ shows maxima near the resonant fields,
where it approaches the equilibrium $\Xone^{\eq}$.
We also display (panel b) the second and third harmonic components
measured at $\w/2\pi=2$\,kHz.
(The high noise-to-signal ratio prevented from obtaining
reliable $\Xnl$ data for $\Hz>1$\,kOe.)
The {\em equilibrium\/} nonlinear susceptibilities, also shown, were
obtained differentiating $\Xone^{\eq}$, measured at the lowest
frequency $\w/2\pi=1$\,Hz.

We clearly see that the magnitudes of the harmonics increase in the
neighborhood of the resonant fields $H_{0}=0$, $H_{1}$ and $H_{2}$,
where states of opposite $S_{z}$ are degenerate and the tunnel
channels open.
Besides, in contrast with the behavior of $\Xone$, both $\Xtwo$ and
$\Xnl$ become, near $H_{0}$, larger than $\Xtwo^{\eq}$ and
$\Xnl^{\eq}$.
Thus, when resonant tunnel sets on, the multi-harmonic response of these
molecular clusters is enhanced.

\begin{widetext} 
In order to understand these results we have derived simple
expressions for the susceptibilities.
This was done by solving, as in Ref.~\cite{gargar2004}, a system of
balance equations for the net population of the two anisotropy
potential wells:
%
\begin{eqnarray}
\label{X1:debye}
\Xone(\w)
&=&
\Xone^{\eq}
\frac{1}{1+\iu\,\w\tau}
\;,
\qquad
\qquad
\Xtwo(2\w)
=
\Xtwo^{\eq}
\frac{1}{1+2\iu\,\w\tau}
-
\Xone^{\eq}
\frac{\iu\,\w\tau'}{(1+\iu\,\w\tau)(1+2\iu\,\w\tau)}
\\
\label{X3:debye}
\Xnl(3\w)
&=&
\Xnl^{\eq}
\frac{1}{1+3\iu\,\w\tau}
-
\Xone^{\eq}
\frac{\frac{1}{2}\iu\,\w\tau''}{(1+\iu\,\w\tau)(1+3\iu\,\w\tau)}
\nonumber\\
& &
{}-
\Xtwo^{\eq}
\frac{2\iu\,\w\tau'}{(1+2\iu\,\w\tau)(1+3\iu\,\w\tau)}
-
\Xone^{\eq}
\frac{2\iu(\w\tau')^{2}}{(1+\iu\,\w\tau)(1+2\iu\,\w\tau)(1+3\iu\,\w\tau)}
\;.
\end{eqnarray}
The equilibrium $\chi_{k}^{\eq}=(\D^{k}M_{z}/\D\Hz^{k})/k!$ are the
derivatives of the magnetization curve, while $\tau$, $\tau'$, and
$\tau''$ are the relaxation time and its corresponding
field-derivatives (all evaluated at the working field $\Hz$).
At $\Hz=0$ we have $\Xtwo^{\eq}\equiv0$ [since $M_{z}(\Hz)=-M_{z}(-\Hz)$]
as well as $\tau'\equiv0$ [from $\tau(\Hz)=\tau(-\Hz)$].
Then $\Xtwo(2\w)\to0$ as $\Hz\to0$, while in $\Xnl$ the last two
terms vanish.
Therefore these equations extend the expressions of
Ref.~\cite{gargar2004} to nonzero bias fields.
\end{widetext}

The first Eq.~(\ref{X1:debye}) gives the ratio $\Xone/\Xone^{\eq}<1$.
Besides $\Xone$ depends, via the product $\w\tau$, on how far the
spins are from thermal equilibrium. 
By contrast, the nonlinear susceptibilities $\Xtwo$ and $\Xnl$ include
also terms depending on $\tau'$ and $\tau''$, i.e., on how sensitive
$\tau$ is to changes of $\Hz$.
As a result, the relaxation time does not simply ``renormalize
frequency'', as occurs with $\Xone$, but it modifies the magnitudes of
the nonlinear responses.
This effect is missed in modelizations of the nonlinear susceptibility
that fail to include the field derivatives of $\tau$ \cite{raietal97}.
As we discuss next, extending our arguments at $\Hz=0$ of
\cite{luietal2004a,lopetal2005}, it is this property that makes the
quantum $\Xtwo$ and $\Xnl$ qualitatively different from the classical
ones.

According to Eq.~(\ref{X1:debye}), the relaxation time of the magnetic
clusters can be estimated from $\Xone$ as $\Xone''/\w\Xone'$ where
$\Xone'$ and $\Xone''$ are the real and imaginary components of the
first harmonic.
This $\tau$ is shown in the inset of Fig.~\ref{fig2}.
The data show minima at the resonant fields, in contrast with the
monotonous behaviour of $\tau$ in classical spins.
At zero field and finite temperatures, Mn$_{12}$ spins are able to
tunnel between those excited magnetic states ($m\sim2$--$4$) for which
such process is not blocked by the internal bias caused by dipolar and
hyperfine interactions \cite{luibarfer98}.
This results in a effective barrier reduced by a few magnetic levels,
say $U\sim\el_{\pm3}-\el_{\pm S}$, so that the thermo-activated
relaxation gets faster ($\delta U\sim4$\,K).
Tunneling is however suppressed as soon as the external bias
$\xi_{m}=2g\mu_{\rm B}m\Hz$ exceeds the tunnel splitting $\Del_{m}$,
slowing down the relaxation (the full barrier has to be overcome).
As a result, $\tau$ is minimum at zero field, whence $\tau''>0$,
while $\tau'$ changes sign from $<0$ to $>0$.

The same features are repeated every time the field brings magnetic
levels again into resonance $H_{n}=n\times(2g\mu_{\rm B})D$.
Therefore, tunneling becomes, at any crossing field, an additional source
of nonlinear response via $\tau'$ and $\tau''$.
Besides, accounting for the signs of the $\tau$ derivatives and
Eqs.~(\ref{X1:debye}) and~(\ref{X3:debye}), one sees that the sign of
the nonlinear susceptibilities can be reversed with respect to the
classical ones.
(In the classical model $\tau$ decreases monotonically with increasing
field \cite{bro63}, inset of Fig.~\ref{fig2}, giving $\tau''<0$ and
$\tau'<0$ for any $\Hz$; the same occurs in a quantum thermoactivation
model not including the possibility of tunneling
\cite{viletal94,zuegar2006}).

It is interesting that both behaviors can be obtained in our case just
varying the external field.
For fields between resonances, tunneling becomes blocked for {\em
all\/} states and the spins reverse by thermal activation over the
total (``classical'') energy barrier.
But when a crossing field is approached, the strong nonlinearity of
$\tau$ shows up with its characteristic contribution to the nonlinear
susceptibilities via $\tau'$ and $\tau''$.

To confirm this interpretation we have computed the nonlinear
responses from Eqs.~(\ref{X1:debye}) and~(\ref{X3:debye}) but
incorporating the relaxation time obtained by solving a Pauli quantum
master equation (as in Ref.~\onlinecite{luibarfer98}).
\begin{figure}
\resizebox{8.5cm}{!}{\includegraphics{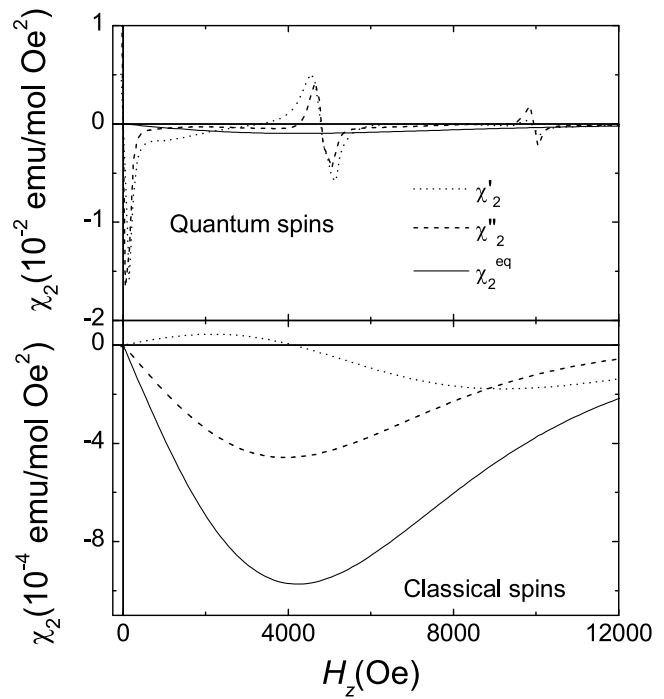}}
\caption{
Theoretical calculations of $\Xtwo(2\w)$ [Eq.~(\ref{X1:debye})] for
quantum and classical spins.
In the latter case, we used Brown's classical formula for $\tau$
\cite{bro63} and in the former $\tau$ was calculated by solving
Pauli's quantum master equation, as described in \cite{luibarfer98}.
Notice the difference in the $\Xtwo$ axis scale (the solid equilibrium
curve is the same in both panels).
}
\label{fig3}
\end{figure}
The results (Fig.~\ref{fig3}) show that the quantum contribution to
$\Xtwo(2\w)$ is dominant near the resonances.
This is due to the smallness of the tunnel splitting of the relevant
states for our Mn$_{12}$ sample: $\Del_{4}\sim2\times10^{-2}$\,K and
$\Del_{2}\sim7\times10^{-1}$\,K.
This means that the fields required to block tunnel via these levels,
albeit relatively small ($\sim20$\,Oe and $1000$\,Oe, respectively),
give rise to relatively large changes in $\tau$, and hence, large
$\tau'$ and $\tau''$.
In the ``classical'' regime, by contrast, the scale of change of
$\tau$ is determined by the anisotropy field $H_{\rm
a}\sim(2S-1)H_{1}$.
As this is very large in Mn$_{12}$ ($\simeq95$\,kOe), one has
a comparatively slow decrease of $\tau$ with $\Hz$.
This correspond to small values of the derivatives $\tau'$ and
$\tau''$ and in turn of the classical (non-tunnel) nonlinear
susceptibilities.

It is also worth mentioning that the tunnel splittings, which
determine the width of the $\tau$ vs $\Hz$ resonances, are further
broadened by dipolar and hyperfine interactions \cite{luibarfer98}.
In fact, the master-equation calculations tell us that tunneling via
lower lying states would give rise to enormous spikes in $\Xtwo$
($\Del_{10}\sim7\times10^{-10}$\,K for the ground state ).
However, these peaks are also very fragile, being easily suppressed by
environmental bias fields and therefore not observed experimentally.
%


\section{
Summary and conclusions
}
\label{conclusions}

We have studied experimentally the nonlinear susceptibilities of a
Mn$_{12}$ acetate molecular magnet in the presence of a
longitudinal field $\Hz$.
The standard method of measuring the harmonics of the response to an
oscillating field $h_{0}\cos(\omega t)$ has been employed.
By using several amplitudes $h_{0}$ we managed to isolate the genuine
nonlinear susceptibilities $\Xtwo$ and $\Xnl$ (oscillating with
$2\omega t$ and $3\omega t$).
The low signal-to-noise ratio (in spite of gluing several single
crystals) prevented from obtaining good $\Xnl$ data; fortunately, we
obtained nice curves for $\Xtwo$, which is the leading nonlinear term
when a bias field is applied.

The analysis and interpretation of the susceptibility curves was done
with help from expressions derived with a simple system of balance
equations (for the potential well populations).
At variance with previous formulae, the field derivatives of the
magnetic relaxation time are captured by our expressions.
This, together with the known strong effect on $\tau(\Hz)$ of resonant
tunneling near the barrier top, permitted to understand the
experimental phenomenology.
We also plugged in those equations the $\tau(\Hz)$ obtained solving a
Pauli quantum master equation for Mn$_{12}$, supporting this
interpretation.

Near the resonant fields $H_{n}$ (matching the levels at both wells)
the $\Xtwo$ vs.\ $\Hz$ curves neatly amplify the resonant tunnel, as
this entails large $\D\tau/\D\Hz$.
For fields in between the $H_{n}$'s, tunnel is blocked and the response
is governed by the thermo-activation over the total barrier, as in the
classical case.
This does not give such large $\tau'$, while its sign is reversed with
respect to the tunnel contribution.
Thus the sensitivity of $\Xtwo$ to the local features of the
$\tau(\Hz)$ curve provides an alternative method to asses if tunneling
plays a role in the relaxation of a superparamagnet; and if so, in
which field ranges it does take place.


\begin{acknowledgments}
Work funded by DGA project {\sc pronanomag} and DGES, projects
MAT02-0166 and BFM2002-00113.
\end{acknowledgments}




\begin{thebibliography}{10}

\item[*]
%
Corresponding author.
email:~fluis@unizar.es.
Tel:~+34976761334;
Fax:~+34976761229.


\bibitem{smm}
D. Gatteschi and R. Sessoli, Angew. Chem. Int. Ed. {\bf 42} (2003) 268;
S.~J. Blundell and F.~L. Pratt, J.~Phys.: Condens. Matter {\bf 16},
R771 (2004); B. Barbara, C. R. Physique {\bf 6} (2005) 934.

\bibitem{leg2002}
A.~J. Leggett, J.~Phys.: Condens. Matter {\bf 14},  R415  (2002).

\bibitem{magtun}
J. L. van Hemmen and A. S\"{u}to, Physica B 141 (1986) 37; M. Enz and R.
Schilling, J. Phys. C {\bf 19} (1986) 1765; D. Garanin, J. Phys. A
{\bf 24} (1991) L61.

\bibitem{zur91}
W.~H. Zurek, Phys. Today {\bf 44},  No.~10, 36  (1991).

\bibitem{kiejoo99}
C. Kiefer and E. Joos,  in {\em Quantum Future}, 
edited by P. Blanchard and A. Jadczyk (Springer, Berlin, 1999), 
Vol.~517 {\rm Lecture notes in physics}, pp.\ 105, quant-ph/9803052.

\bibitem{frietal96}
J.~R. Friedman, M.~P. Sarachik, J. Tejeda, and R. Ziolo,
Phys. Rev. Lett. {\bf  76},  3830  (1996).

\bibitem{heratal96}
J.~M. Hern{\'a}ndez, X.~X. Zhang, F. Luis, J. Bartolom{\'e}, 
J.~Tejada, and R.  Ziolo,
Europhys. Lett. {\bf 35},  301  (1996).

\bibitem{thoetal96}
L. Thomas, F. Lionti, R. Ballou, D. Gatteschi, R. Sessoli, and B. Barbara,
  Nature {\bf 383},  145  (1996).

\bibitem{wer2001}
W. Wernsdorfer, Adv. Chem. Phys., {\bf 118}, 99 (2001), cond-mat/0101104.

\bibitem{luietal2004a}
F. Luis, V. Gonz{\'{a}}lez,
A. Mill{\'{a}}n, and J.~L. Garc{\'{\i}}a-Palacios,
Phys. Rev. Lett. {\bf 92},  107201  (2004).

\bibitem{lopetal2005}
R. L{\'o}pez-Ruiz and F. Luis and V. Gonz{\'{a}}lez 
and A. Mill{\'{a}}n and J. L. Garc{\'{\i}}a-Palacios,
Phys. Rev. B {\bf 72} (2005) 224433.

\bibitem{garsve2000}
J.~L. Garc{\'{\i}}a-Palacios and P. Svedlindh,
Phys. Rev. Lett. {\bf 85},  3724   (2000).

\bibitem{gargar2004}
J.~L. Garc{\'{\i}}a-Palacios and D.~A. Garanin,
Phys. Rev. B {\bf 70}, 064415  (2004).

\bibitem{bro63}
W.~F. Brown, Jr., Phys. Rev. {\bf 130},  1677  (1963).

\bibitem{raietal97}
Y.~L. Raiker, V.~I. Stephanov, A.~N. Grigorenko
and P.~I. Nikitin, Phys.  Rev. E {\bf 56}, 6400 (1997).

\bibitem{luibarfer98}
F. Luis, J. Bartolom{\'{e}}, and J.~F. Fern{\'{a}}ndez,
Phys. Rev. B {\bf 57},  505  (1998).

\bibitem{viletal94}
J.~Villain, F.~Hartmann-Boutron, R.~Sessoli, and A.~Rettori,
Europhys. Lett. {\bf 27}, 159 (1994).

\bibitem{zuegar2006}
D. Zueco and J.~L. Garc{\'{\i}}a-Palacios,
Phys. Rev. B {\bf 73}, 104448 (2006), cond-mat/0509627.

\end{thebibliography}
\end{document}